\documentclass[aps,prl,showpacs,twocolumn,floats]{revtex4}
\usepackage{graphicx}
\usepackage{amssymb}
\usepackage{amsbsy}
\usepackage{amsmath}

\begin{document}
\def\be{\begin{equation}} 
\def\ee{\end{equation}}
\def\bearr{\begin{eqnarray}}
\def\eearr{\end{eqnarray}}
\def\tc{$T_c~$}
\def\tcl{$T_c^{1*}~$}
\def\c2{ CuO$_2~$}
\def\lsco{LSCO~}
\def\half{$\frac{1}{2}$~}
\def\thalf{$\frac{3}{2}$~}
\def\kfese{$\rm K_2Fe_4Se_5$~}
\def\f4{$\rm Fe_4$~}
\def\7half{$\rm 7\frac{1}{2}$~}

\title{Possibility of Skyrmion Superconductivity in Doped Antiferromagnet 
K$_2$Fe$_4$Se$_5$}

\author{ G. Baskaran}
\affiliation
{Quantum Science Centre, The Institute of Mathematical Sciences, C I T Campus,
Chennai 600 113, India\\
Perimeter Institute for Theoretical Physics, 31 Caroline St N, Waterloo,
Ontario, Canada N2L 2Y5}

\begin{abstract}
Intercalated Fe chalcogenide K$_2$Fe$_4$Se$_5$ family exhibits high Tc ($\sim$ 30 K) 
superconductivity
and spin-8 high T$_{\rm N}$ ($\sim$ 560 K) antiferromagnetism (AFM). We present a model
Hamiltonian and suggest \textit{Skyrmion superconductivity}. A doped electron
creates an orbitally non degenerate S=7\half state in a F$_4$Se$_9$ cluster and
moves in a single correlated band, exchange coupled to a robust S=8 AFM order.
Skyrmion, a topological excitation of 2 dimensional AFM order acquires an
\textit{induced charge -2e} through a quantum anomaly and becomes a Cooper pair.
Superconductivity emerges for a range of doping. Fluctuating superconductivity,
arising from preformed Cooper pairs (stable Skyrmions), around room temperatures 
is likely. There is hope for higher Tc's in other high spin Mott insulators. 
\end{abstract}

\maketitle

Discovery of superconductivity in Fe based pnictides \cite{1hasano,2chinese}
and chalogenides \cite{3Wu} have created another wave of excitement in condensed
matter physics. Even though Tc has not matched the scale of 100 K seen in cuprates, 
Fe systems continue to offer surprises. A recent surprise is
the discovery \cite{5Bao} of superconductivity in intercalated Fe chalogenide
family, where a robust antiferromagnetic order with a sublattice moment close to
13 $\mu_{\rm B}$ and a T$_N \approx $ 560 K seems to coexist with a high Tc
superconductivity of Tc $\approx$ 30 K. From materials science point of view the
vacancy order present in \kfese has enabled a rich physics. Ordered vacancies
is a new route to novel states of quantum matter \cite{AkbarDiamond}. 

In the present article we provide a minimal model and identify a mechanism of
electron pairing. We use some general considerations and present certain
consequences that are compelling. We show that a single band model of strongly
correlated electrons, in a square lattice, close to half filling exchange
coupled to a square lattice of large spin-7\half core moments of \f4 plaquettes
captures the essential low energy physics. In our theory topological excitations
of an ordered 2 dimensional Heisenberg antiferromagnet, namely 
Skyrmion \cite{Skyrmion}, plays a
fundamental role. Relevance of topological excitations \cite{DPW,6MeronPWA} to
lightly doped quantum antiferromangets began in 1987 with the advent of RVB
theory of high Tc superconductivity. This study continues \cite{SkyrmionOthers,DungHai}. 

Briefly, kinetic energy frustration of doped electrons in the presence of a rigid
AFM order is relieved by a Skyrmion which help delocalize 2 electrons, using its spin twist. 
Skyrmion, through a quantum anomaly, acquires an \textit{induces} charge -2e and
becomes a Cooper pair. For very low doping we get an ordered
Skyrmion crystal. Beyond a range of doping Skyrmion crystal quantum melts,
leading to \textit{Skyrmion superconductivity}. For even higher dopings
ferromagnetism is likely to intervene, with a possible island of p-wave
superconductivity (figure 4).

Following the discovery of superconductivity in FeSe \cite{3Wu},
a new family, A$_x$Fe$_y$Se, with intercalated alkali atoms (A = K,Cs,Rb) was
synthesized \cite{5Bao}. Stoichiometric \kfese has ordered Fe vacancies 
and $\sqrt{5} \times \sqrt{5}$ periodicity. Experiments indicate 
\cite{Neutron,ExptMottInsulator} that \kfese is a Mott insulator. Neutron 
scattering\cite{Neutron} shows that the \f4 plaquettes attain a large cluster 
moment and reaches a S=8 spin state. Further the
plaquettes develop a $(\pi,\pi)$ long range AFM order (figure 1) with a 
T$_{\rm N} \sim$ 560 K.


\begin{figure}[t]
\begin{center}
\includegraphics[width=2.8in]{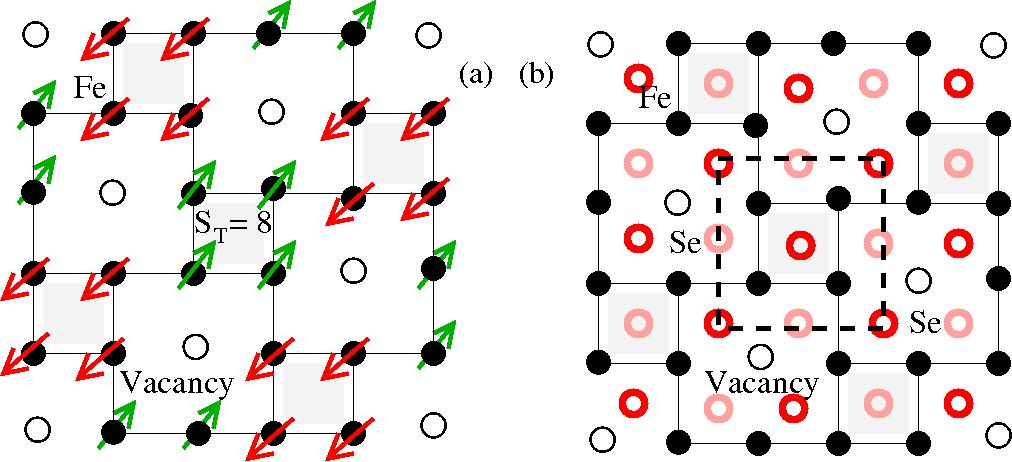}
\caption{$\sqrt{5}\times\sqrt{5}$ a) vacancy (o) order and S=8  (=2 $\times$ 4)
AFM order 
of \f4 plaquettes in  FeSe layer of \kfese. Only Fe atoms (\textbullet) are shown. 
Fe$^{2+}$ is in S = 2, the maximal spin state.
b) Se atoms above (red) and below (pink) Fe layers are shown. Dashed square in the 
middle marks a Fe$_4$Se$_9$ unit.}
\end{center}
\end{figure}

Superconducting Tc is around 30 K, in various members of the family. Many
experiments show coexistence of magnetism and superconductivity
\cite{Neutron,ExptMottInsulator,exptCoexistence}. However, NMR and STM results \cite{11NMR} 
seem to challenge coexistence.

There are some theoretical studies on this system \cite{MagnetismTheory}. We
present a new approach and develop a minimal microscopic model to understand
magnetic and superconducting properties. We first propose that \textit{\f4 plaquette is
a basic electronic building block}. \textit{Electronic integrity} of \f4
plaquette is revealed by its robust total spin-8 state arising from strongly
exchange coupled four Fe atoms that carry spin-2 each. The S=8 plaquette spins
have a $(\pi,\pi)$ AFM order in the 2D square lattice of plaquettes. An
important question is where do doped carriers go ? This question is similar to
Zhang-Rice singlet formation in cuprates. In view of available experiments 
we will focus on electron doping. 

Simple quantum chemical and phenomenological considerations suggest that a doped
electron enters a \f4 plaquette and create an \textit{orbitally
non degenerate} spin-7\half, \textit{hexadectic~state} in the \f4 plaquette \cite{PWA}.
Briefly, Fe atoms of \f4 plaquette are tetrahedrally coordinated by Se atoms and
form a Fe$_{4}$Se$_{9}$ unit. In view of strong hybridization of 3d orbitals
of Fe and 4s/4p orbitals of Se atoms, the 16 electrons that makeup S=8 spin
state of \f4 plaquette reside in 16 \textit{plaquette orbitals} in a strongly
exchange coupled parallel spin state (figure 1b). A doped electron enters the
top most singly occupied non degenerate orbital. We call this as \textit{valence
orbital} of \f4 plaquette that carries a \textit{spin-\half valence moment}.

\begin{figure}[t]
\begin{center}
\includegraphics[width=2.5in]{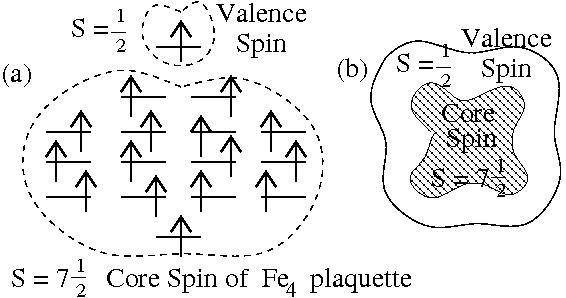}
\caption{a) Schematic energy levels of 16 singly occupied and spin polarized
plaquette orbitals
of the Fe$_4$S$e_{9}$ cluster. b) The strongly hybridized core spin-7\half
moment and the valence 
spin-\half moment distributed in the \f4 plaquette is schematically shown.}
\label{EnergyLevels}
\end{center}
\end{figure}

Anticipating addition of an electron (doping) we conveniently separate S=8 spin
of \f4 into S=7\half core spin moment $\textbf{S}_i$, ferromagnetically coupled
by a \textit{plaquette Hund coupling} J$_{\rm H}$ to a S=\half valence moment 
$\textbf{s}_i$
(figure 1b). Neighboring \f4 plaquette spins are antiferromagnetically coupled
by an effective superexchange J. Thus the spin Hamiltonian of the Mott insulator
is:
\bearr
H_{\rm AFM} = - J_{\rm H} \sum_{i} \textbf{S}_{i} \cdotp \textbf{s}_i +
J\sum_{\langle ij \rangle}  (\textbf{S}_{i} + \textbf{s}_i) \cdotp
(\textbf{S}_{j} &+& \textbf{s})\\
= J\sum_{\langle ij \rangle}  \textbf{S}_{i} \cdotp \textbf{S}_{j}
- J_{\rm H} \sum_{i} \textbf{S}_{i} \cdot \textbf{s}_i +
\frac{J}{2}\sum_{\langle ij \rangle}  (\textbf{S}_{i} \cdot \textbf{s}_j &+&
\textbf{S}_{j} \cdot \textbf{s}_i)\nonumber
\eearr
Here $ \textbf{s}_i \equiv c^{\dagger}_{i\alpha} \vec{\sigma}_{\alpha\beta}
c^{}_{i\beta}$, is the valence spin-\half operator written interms of electron
operators ($c^{\dagger}_{i\sigma}, c^{}_{i\sigma}$) of the \f4 plaquette valence
orbital. For simplicity we ignore the weak c-axis exchange coupling and any
anisotropic spin interaction terms.

In the limit J$_{\rm H} \gg $ J, equation 1 becomes a Spin-8 Heisenberg AFM 
Hamiltonian on a
square lattice. The effective exchange coupling J can be estimated from the
\textit{acoustic spin wave} band width from inelastic neutron scattering
results\cite{Neutron}. The estimate is JS = 8J $\approx$ 17 meV. Location of the
first \textit{optic spin wave} branch give a value for J$_{\rm H} \approx$ 110
meV.

The Hamiltonian of our electron doped system is a single band model of plaquette
valence spin electrons
\textit{close to half filling}, coupled to S = 7\half core spins of \f4
plaquettes:
\bearr
H &=&  - \sum_{ij} t_{ij} c^{\dagger}_{i\sigma}c^{}_{j\sigma} + {\rm h.c} +
J\sum_{\langle ij \rangle} \textbf{s}_i \cdot \textbf{s}_j +
J\sum_{\langle ij \rangle}  \textbf{S}_{i} \cdotp \textbf{S}_{j} \nonumber \\
&-& J_{\rm H} \sum_{i} \textbf{S}_{i} \cdot \textbf{s}_i +
\frac{J}{2}\sum_{\langle ij \rangle}  (\textbf{S}_{i} \cdot \textbf{s}_j +
\textbf{S}_{j} \cdot \textbf{s}_i) 
\eearr\
with a local constraint $\sum_{\sigma} c^{\dagger}_{i\sigma}c^{}_{i\sigma} \neq
0 $ (for electron doping). Here t is the inter plaquette hopping matrix element. 
From known band
structure results, after folding the BZ due to vacancy ordering, we estimate t
$\sim$ 50 meV. The band width W $\sim$ 400 meV is large compared to the
charging energy U$_{\rm eff} \approx \frac{e^2}{\rm \epsilon_0 R} \sim $ 500 meV
of the (Fe$_{2}$Se$_{5})^{2+}$ plaquette. Here R is the size of the \f4
plaquette. Our doping fraction x is with reference to the \f4 plaquette orbital
band. Thus in the formula ${\rm K_{2+y}Fe_{4}}Se_{5}$, our doping x =
$\dfrac{y}{4}$.

The low energy effective Hamiltonian of the interacting core spin Heisenberg
antiferromagnet is a 2 + 1 dimensional O(3) nonlinear sigma model with action:
\be
S\sim\int dx~dy~d\tau [(\partial_\mu {\textbf{n}(\textbf{r}))}^2 
- \frac{1}{v_s^2}(\partial_t {\textbf{n}(\textbf{r}))}^2]
\ee
Here $\mu$ = x,y; $v_{s}$ is the spin wave velocity and $\textbf{n}(\textbf{r})$ is the
normalized, $\textbf{n}(\textbf{r})
\cdot \textbf{n}(\textbf{r}) = 1$,  sublattice magnetization. 

The above system supports small amplitude gapless spin wave (Goldstone mode)
excitations and finite energy topological excitations namely Skyrmions.
Skyrmions are analogue of a pair of domain walls of 1D Heisenberg
antiferromagnet. In 1D system a single domain wall is a $\pi$$-twist$ that
interpolates up and down spin vacua in a continuous fashion. In the process it
effectively decouples a single spin (a \textit{spinon} localized around the domain wall) 
from the antiferromagnetic mean field. In the language of spin density wave (similar to
polyacetylene) it creates a midgap state containing an unpaired spin.

\begin{figure}[t]
\begin{center}
\includegraphics[scale=0.3]{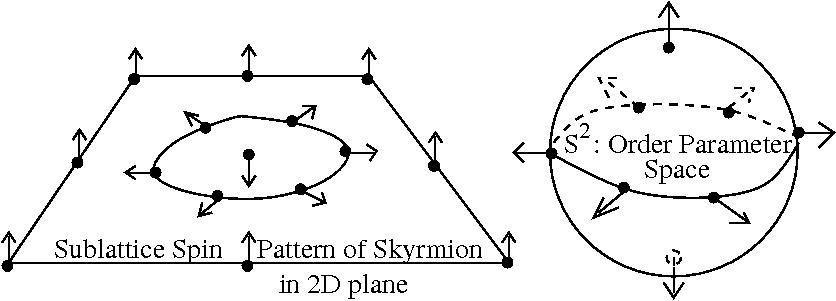}
\caption{Spin configuration of a Skyrmion. A circle in the middle has only xy
component of sublattice spins and defines an XY-vortex. An anti Skyrmion will
have an antivortex.}
\label{Skyrmion}
\end{center}
\end{figure}

A Skyrmion interpolates the down spin vacuum in its central region and up spin
vacuum at the boundary, in a radially symmetric fashion (figure 3). In the process it
effectively decouples two spin-\half\textit{ spinons} \cite{6MeronPWA}. This can be seen in several
ways\cite{6MeronPWA,7John,DPW}. A simple way is in the SDW band picture,
as studied by John and collaborators\cite{7John}. A skyrmion or 2
half skyrmions (2 merons) configuration traps two unpaired spins in two mid gap
states. 

Skyrmion has topological properties. It carries chirality or Pontryagin index Q 
(= 0, $\pm$ 1, $\pm$ 2, ...). Its sign measures the sense, and magnitude number 
of windings, when we
map order parameter $\textbf{n}(\textbf{r})$ field from the 2D plane 
onto S$^{2}$, the surface of an ordinary sphere:
\be
Q \equiv \frac{1}{8\pi} \int dx~dy ~ \textbf{n}\cdot(\partial_x\textbf{n} \times
\partial_y\textbf{n})
\ee
In the language of homotopy theory, $\rm{\varPi_2(S^2) = Z}$. Skyrmion has 
a zero mode associated with a residual U(1) symmetry about the z-axis of the 
spin configuration. Quantized zero mode represents spin current fluctuation 
\cite{GB.Dyon} and quantum Skyrmion is a \textit{chiral doublet}. This is a 2D 
analogue of Affleck's  dyon\cite{Affleck.Dyon} in 1D Heisenberg antiferromagnets.

Valence electrons living in a single band in our model are exchange coupled to
an antiferromagnetic mean field provided by the core spin moments from \f4
plaqauettes. An electron in a rigid AFM background can not delocalize. We have 
kinetic energy frustration. Skyrmion (Q = $\pm$ 1) relieves the frustration 
by delocalizing 2 electrons using spin current (spin twist) under its wings.

Taking care of the local constraints and exchange interactions in a mean field
fashion we get the mean field Hamiltonian of the valence electrons system as:
\be
H_{\rm mF} \approx  \sum_{ij} {\tilde{t}}_{ij}
c^{\dagger}_{i\sigma}c^{}_{j\sigma} + {\rm h.c} - 
{\tilde{J}}_{\rm H}S\sum_i \textbf{n}_i\cdot c^{\dagger}_{i\alpha}
\vec{\sigma}_{\alpha\beta} c^{}_{i\beta}
\ee
Here $\tilde{t}_{ij}$ is a renormalized interplaquette hopping matrix element,
(we choose $\pi$ flux mean field solution); $\tilde{J}_H = J_H + 4 J$. The
$(\pi,\pi)$ antiferromagnetic order introduces a gap in the two Dirac nodes.
To calculate induced electric charge, in the presence of a slowly varying order
parameter \textbf{n}(\textbf{r}) and U(1) gauge field A$^c_\mu$, we use the
language of 2D Dirac fermions and follow Wigmann and Abanov\cite{DPW}.
The effective action of the valence electron quasi particles coupled to slowly
varying background AFM field is:
\be
S \sim \int d^3x ~ {\bar\psi}~[-i\gamma_{\mu}(\partial_\mu + A_\mu^c) +
im~\vec{\sigma}\cdot \textbf{n}({\textbf r})]~\psi
\ee
Here $\mu$ = x,y and $\tau$, $\bar\psi = i\psi^\dagger \gamma^0$, and
$\gamma^0,\gamma^x,\gamma^y \equiv \sigma^z,\sigma^y,-\sigma^x$. We have
rescaled the Fermi velocity of electron to 1 for brevity. Further m $\approx
SJ_{\rm H}$ is a \textit{mass gap} provided by the strong exchange field.
Integrating out the fermions and in the large mass approximation one gets
\cite{DPW}:
\be
S = S_{\rm R} + 2i A_\mu^c J_\mu^T
\ee
Here $J_\mu^T(\textbf{r}) = \frac{1}{8\pi} \epsilon^{\mu\nu\lambda}
\textbf{n}\cdot\partial_\nu \textbf{n} \times \partial_\lambda \textbf{n}$ is
the topological spin current density, whose time component $J_0^T(\textbf{r})
\equiv \frac{1}{8\pi} \textbf{n}\cdot(\partial_x\textbf{n} \times
\partial_y\textbf{n})$ is the Skyrmion density (equation 4).
The real part of the induced interaction renormalizes the NLS action (equation
3). The imaginary part is the key U(1) gauge field $A^c_\mu$ - topological spin
current  $J_\mu^T$ coupling. The time component of the above, $2i A_0^c J_0^T$
is the coupling of dopant chemical potential to Skyrmion density (equation 4).
Hence we get an induced U(1) (electric) charge $Q_{\rm Skyrmion}^c$ given by the
expression\cite{DPW}:
\be
Q_{\rm Skyrmion}^c = \pm ~2e ~|Q|
\ee
where Q is the Pontryagin index or Skyrmion number. 

To calculate the spin of the - 2e Skyrmion, we start with an even number of
electrons in an ordered antiferromagnet and create one Skyrmion. This generates
two singly occupied mid gap states. Addition of two electrons to these midgap
states completely fill the two states and create spin zero midgap states. Thus
nominally a 2D Skyrmion carrying charge - 2e is a spin singlet. 

2e Skyrmion, a charged chiral doublet, as discussed by Lee \cite{DungHai} for
the case of a spin liquid, represents a L$_z$ = $\pm$ 2 angular momentum, 
d $\pm$ id state of a Cooper pair. Skyrmions that are real combination of the 
above doublet will lead to a \textit{real} d-wave superconducting state. 
Otherwise we will get a d $\pm$ id superconducting state that breaks 
PT symmetry. This will be a chiral superconductor $\acute{a}$ la Laughlin. 

In an ideal O(3) non linear sigma model in 2D, Skyrmion  energy is finite and
size independent. In practice, coulomb repulsion among 2 electrons, spin anisotropy 
and interchain coupling determine the size. For our case, the mean size of Skyrmion 
is determined by the inter Skyrmion distance. Cooper pair size or coherence length is
selfconsistantly limited by mean inter Cooper pair distance. 

Skyrmion delocalization and kinetic energy gain is determined by their mutual
overlap, proportional to x. As a first approximation, Skyrmion band width is a
fraction x of the fermi energy Wx of free dopants. That is, the Skyrmion band
width W$_{\rm Skyrmion} \sim $ Wx$^2$. Since W $\sim$ 400 meV, Wx$^{2} \sim$ 4
meV, for a doping of x $\sim$ 0.1. As -2e Skyrmions are Bosons this is also Bose
condensation energy scale. This energy is in the right ball park of the observed
Tc $\approx$ 30 K. Our superconductor is best viewed as a 2D Kosterlitz-Thouless
condensed state of interacting bosons.
Weak Josephson coupling between layers will make it a 3D superconductor. 

Skyrmions are stable, in principle upto Neel temperature, as long as AFM order
survives. In this sense 2e Skyrmions are \textit{preformed Cooper pairs} in the
background of an AFM order. As a consequence we could see Cooper pairs well
above Tc, through for example anomalous Nernst effect. We predict that \kfese
family offers a wonderful and some what unique opportunity to explore and
utilise fluctuating superconductivity around room temperature scales.

In our theory the large spin AFM order and superconductivity coexist for a range
of doping. Dopant delocalization will eventually destroy the long range AFM order.
\begin{figure}[t]
\begin{center}
\includegraphics[width=2.2in]{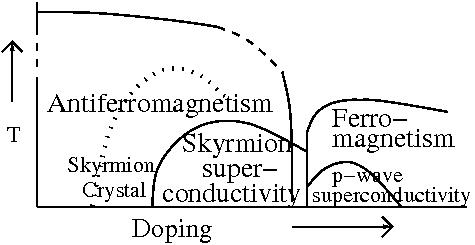}
\caption{Schematic phase diagram for doped \kfese family.}
\label{PhaseDiagram}
\end{center}
\end{figure}
In the low doping regime unscreened coulomb interaction will localize Skyrmions
resulting in Skyrmion crystal or Skyrmion glass. Berciu and John \cite{7John}
have  discussed this for cuprates. A universal resistivity maxima seen in
\kfese superconductor could indicate onset of Skyrmion crystallization 
frustrated by growing quantum fluctuations (dashed line in figure 4 inside AFM dome).

It is known that large spin doped antiferromagnets leads to ferromagnetism through 
\textit{double exchange}, for example in doped 
manganites, a spin-2 Mott antiferromagnet. As shown by de Gennes, small doping
leads to spin twist (canting), before ferromagnetic order sets in at higher 
dopings. In a 2 dimensional system such as ours, the system naturally organize 
spin twists in the form of topologically stable Skyrmion eigen excitations and 
help delocalization.  What is exciting is that \textit{superconductivity 
is an alternative to ferromagnetism in 2 dimensions}. So we have a new avenue for 
high Tc search, namely high spin antiferromagnets in lower dimensions.

It also follows from double exchange physics that heavily doped \kfese will 
eventually become a ferromagnetic 
metal. If we take clue from Sr$_2$RuO$_4$, another 2D even electron (Ru$^{2+} =
4d^4$) system, Hund coupling and a growing ferromagnetic fluctuation could 
stabilise an island of spin triplet p-wave superconductivity.

A robust AFM magnetic order of a large spin Mott antiferromagnet offers
exciting possibilities, as shown schematically in figure 4. It will be useful
to study doped \kfese family experimentally with controlled doping.

I thank D. Abanin, P.W. Anderson, T. Imai, D.H. Lee, N.P. Ong and T. Xiang for
discussion. This research was supported by Perimeter Institute for Theoretical
Physics.

\end{document}